\documentclass[a4paper]{jpconf}

\usepackage{graphicx}
\usepackage{amssymb}
\usepackage{amsmath}
\usepackage{iopams}

\begin{document}

\title{Behavior of universal critical parameters in the QCD phase diagram} 

\author{Marcus~Bluhm$^{\,1}$, Marlene~Nahrgang$^{\,2,3}$, Steffen~A.~Bass$^{\,3}$ and Thomas~Sch\"afer$^{\,1}$}
\address{$^1$ Department of Physics, North Carolina State University, Raleigh, NC 27695, USA}
\address{$^2$ SUBATECH, UMR 6457, Universit\'{e} de Nantes, \'{E}cole des Mines de Nantes, IN2P3/CNRS. 
4 rue Alfred Kastler, 44307 Nantes cedex 3, France}
\address{$^3$ Department of Physics, Duke University, Durham, NC 27708, USA}

\ead{mbluhm@ncsu.edu}

\begin{abstract}
 We determine the dependence of important parameters for critical fluctuations on temperature and baryon chemical 
 potential in the QCD phase diagram. The analysis is based on an identification of the fluctuations of the order 
 parameter obtained from the Ising model equation of state and the Ginzburg-Landau effective potential approach. 
 The impact of the mapping from Ising model variables to QCD thermodynamics is discussed. 
\end{abstract}

\section{\label{sec:1} Introduction}

Effective models for QCD thermodynamics~\cite{Asakawa:1989bq,Berges:1998rc,Halasz:1998qr} predict that the chiral 
phase transition in QCD is of first order at large baryon chemical potential $\mu_B$. At $\mu_B=0$, lattice QCD 
simulations established that the transformation between the denser deconfined quark matter phase at high 
temperature $T$ and the low-$T$ hadronic matter phase is an analytic crossover~\cite{Aoki:2006we}. As a 
consequence, the first-order phase transition line should terminate in a second-order chiral critical point at 
some non-zero $\mu_B$. A critical point is characterized through the growth of the fluctuations of the order 
parameter $\sigma$ in the scaling region around it. In the thermodynamic limit, these fluctuations diverge at the 
critical point with a given exponent of the diverging correlation length $\xi$. 

Experimentally, the phases of QCD matter are studied in heavy-ion collision experiments by varying the beam energy 
and the system size. The quest for finding the conjectured critical point in the QCD phase diagram has triggered 
an extensive effort culminating in the recently completed beam energy scan phase~I at 
RHIC~\cite{Adamczyk:2013dal,Adamczyk:2014fia,Adare:2015aqk} and being continued with the NA61 experiment at 
CERN-SPS and the phase~II of RHIC's beam energy scan program in the future. Signaling the presence of a critical 
point, large event-by-event fluctuations~\cite{Stephanov:1998dy,Stephanov:1999zu} of conserved quantities such as 
electric charge or baryon number have been predicted. In reality, the actual growth of the correlation length is 
limited by the finite size and, more importantly, lifetime of the created system. Thus, higher-order non-Gaussian 
cumulants of the event-by-event distributions, which depend more strongly on the growth of 
$\xi$~\cite{Stephanov:2008qz}, are particularly interesting. These exhibit certain patterns in the QCD phase 
diagram~\cite{Asakawa:2009aj,Stephanov:2011pb} which are governed by the qualitative behavior of universal 
critical parameters. 

Assuming that QCD belongs to the same universality class as the three-dimensional Ising 
model~\cite{Berges:1998rc,Halasz:1998qr} allows one to identify the expectation value of the order parameter of 
the chiral phase transition with the order parameter in the spin model, the magnetization $M$. Fluctuations of 
the order parameter in the scaling region can then be determined from the critical equation of state for $M$. This 
enables us to study the behavior of the universal critical parameters in this work. A similar approach was 
previously used in~\cite{Nonaka:2004pg,Bluhm:2006av} to construct an equation of state with critical point for 
QCD matter. 

In the following, we employ the linear parametric representation~\cite{Schofield:1969zza} of the critical equation 
of state for the magnetization with critical exponents $\beta=1/3$ and $\delta=5$ 
\begin{align}
 \label{equ:1}
 M & = M_0 R^\beta \theta \,,\\
 \label{equ:2}
 r & = R (1-\theta^2) \,,\\
 \label{equ:3}
 h & = R^{\beta\delta} \tilde{h}(\theta) \,.
\end{align}
In this representation $M(r,h)$, which is a function of reduced temperature $r=(T-T_c)/T_c$ with spin model 
critical temperature $T_c$ and of reduced external magnetic field $h=H/H_0$, is expressed in terms of the 
auxiliary variables $R\geq 0$ and $\theta$ as $M(R,\theta)$. The parametrization is uniquely defined within 
$-\theta_0\leq\theta\leq\theta_0$, where for $\tilde{h}(\theta)=3\theta(1-2\theta^2/3)$, which is an odd function 
of $\theta$, one finds $\theta_0=\sqrt{3/2}$ as the relevant non-trivial root. $M_0$ and $H_0$ are normalization 
constants of mass dimension one and three, respectively. 

In spin model coordinates, the critical point is located at $r=h=0$ ($R=0$). At $h=0$, one finds a first-order 
phase transition for $r<0$ ($R>0$ and $\theta=\pm\theta_0$), while for $r>0$ there is a crossover ($R>0$ and 
$\theta=0$). The behavior of the order parameter in the scaling region is such that 
$M(r=0,h)\sim|h|^{1/\delta} sgn(h)$ and $M(r,h\to 0^+)\sim|r|^\beta$ for $r<0$, which can be assured by imposing 
conditions on $M_0$ and $H_0$ such that $M$ is positive for $h\geq 0^+$~\cite{Nonaka:2004pg}. 

\section{\label{sec:2} Order parameter fluctuations}

The equilibrium cumulants quantifying the fluctuations of the order parameter in the scaling region are determined 
from derivatives 
\begin{equation}
\label{equ:4}
 \langle(\delta\sigma)^n\rangle_c = 
 \left(\frac{T}{VH_0}\right)^{n-1}\left.\frac{\partial^{n-1}M}{\partial h^{n-1}}\right|_r 
\end{equation}
of $M$ with respect to the external magnetic field at fixed $r$. Explicit expressions in terms of $R$ and $\theta$ 
based on the linear parametric representation can be found in~\cite{Mukherjee:2015swa}. For given $r$ and $h$, 
these cumulants can then be determined by making use of Eqs.~(\ref{equ:2}) and~(\ref{equ:3}). 

The cumulants of the fluctuations of the order parameter can, likewise, be determined from the corresponding 
probability distribution which depends on an effective action for the order parameter of Ginzburg-Landau 
type~\cite{Mukherjee:2015swa,Stephanov:2008qz}. Including up to quartic interaction terms one finds 
\begin{align}
 \label{equ:5}
 \langle (\delta\sigma)^2 \rangle & = \frac{T}{V}\xi^2 \,,\\
 \label{equ:6}
 \langle (\delta\sigma)^3 \rangle & = -2\lambda_3\frac{T^2}{V^2}\xi^6 \,,\\
 \label{equ:7}
 \langle (\delta\sigma)^4 \rangle_c & = 6\left(2(\lambda_3\xi)^2-\lambda_4\right)\frac{T^3}{V^3}\xi^8 
 \,.
\end{align}
In the scaling region, the cubic and quartic interaction strengths depend on the correlation length as 
$\lambda_3=\tilde{\lambda}_3\,T(T\xi)^{-3/2}$ and $\lambda_4=\tilde{\lambda}_4(T\xi)^{-1}$, where we neglected 
small anomalous scaling dimension corrections in the cumulant expressions. 

Strictly speaking, the expressions in Eqs.~(\ref{equ:5})~-~(\ref{equ:7}) are only valid in the scaling region as 
long as the correlation length is small compared to the macroscopic length scale of the considered system. Keeping 
this limitation in mind we can, nonetheless, identify the expressions for the cumulants following from 
Eq.~(\ref{equ:4}) with those in Eqs.~(\ref{equ:5})~-~(\ref{equ:7}). This yields parametric expressions for $\xi$, 
$\tilde{\lambda}_3$ and $\tilde{\lambda}_4$ in terms of the auxiliary spin model variables reading 
\begin{align}
\label{equ:8}
 \xi & = \sqrt{\frac{M_0}{H_0}} \frac{1}{R^{2/3}(3+2\theta^2)^{1/2}} \,, \\
\label{equ:9}
 \tilde{\lambda}_3 & = \frac{2\theta(9+\theta^2)}{(3-\theta^2)(3+2\theta^2)^{3/4}} C \,,\\
\label{equ:10}
 \tilde{\lambda}_4 & = 2 \tilde{\lambda}_3^2 + 
 2 \frac{(81-783\theta^2+105\theta^4-5\theta^6+2\theta^8)}{(3-\theta^2)^3(3+2\theta^2)^{3/2}} C^2 \,,
\end{align}
where $C=(T^2H_0/M_0^5)^{1/4}$. The parameters $\tilde{\lambda}_3$ and $\tilde{\lambda}_4$ are found to be 
dimensionless and independent of $R$ which implies that they remain finite in the entire domain of the parametric 
representation, while $\xi$ is of inverse mass dimension and diverges as $R\to 0$. 

According to Eq.~(\ref{equ:9}), $\tilde{\lambda}_3$ is an odd function of $h$ which is negative for $h<0$. At 
$h=0$, one finds $\tilde{\lambda}_3=0$ for $r>0$, while for $r<0$ the dimensionless parameter approaches 
$\tilde{\lambda}_3\to\pm 7\,C/6^{1/4}$ as $h\to 0^{\pm}$. The qualitative behavior of the fourth-order cumulant 
$\langle (\delta\sigma)^4 \rangle_c$ is determined by the behavior of the parameter difference 
$2 \tilde{\lambda}_3^2 - \tilde{\lambda}_4$. As evident from Eq.~(\ref{equ:10}), this difference is an even 
function of $h$, which is positive for all $r<0$ with maximum value 
$(2 \tilde{\lambda}_3^2 - \tilde{\lambda}_4)=128\,C^2/27$ at $h=0$. For $r>0$, instead, one finds an interval 
$-h_0\leq h\leq h_0$ in which the parameter difference becomes negative with minimum value 
$(2 \tilde{\lambda}_3^2 - \tilde{\lambda}_4)=-4\,C^2/9$ at $h=0$. The size of this interval depends on the 
value of r and shrinks to zero, $h_0\to 0$, as $r\to 0^+$. 

\section{\label{sec:3} Parameter behavior in the QCD phase diagram}

The parameters $\tilde{\lambda}_3$ and $\tilde{\lambda}_4$ in Eqs.~(\ref{equ:9}) and~(\ref{equ:10}) are universal 
functions of the spin model variables in the scaling region. However, the mapping from $r$ and $h$ to $\mu_B$ and 
$T$ in QCD thermodynamics is not universal but strongly model-dependent. By relating the density difference from 
the critical density in QCD to the magnetization $M$, this mapping has to satisfy, nonetheless, certain 
constraints near the critical point based on universality class arguments. The QCD critical point with baryon 
chemical potential $\mu_B^{\textrm{cp}}$ and temperature $T^{\textrm{cp}}$ must be located at $r=h=0$, positive 
$r$-values must correspond to the QCD crossover regime and positive $h$-values have to be realized in the denser 
phase. 

In practice, these conditions can be assured by a simple linear mapping employing as auxiliary variables 
\begin{equation}
\label{equ:11}
 \tilde{r} = \frac{(\mu_B-\mu_B^{\textrm{cp}})}{\Delta \mu_B^{\textrm{cp}}} \quad , \quad 
 \tilde{h} = \frac{(T-T^{\textrm{cp}})}{\Delta T^{\textrm{cp}}} \,.
\end{equation}
The parameters $\Delta T^{\textrm{cp}}$ and $\Delta \mu_B^{\textrm{cp}}$ relate scales in the spin model 
coordinate system with the unknown size of the critical region in QCD. As by definition $\tilde{r}>0$ in the QCD 
first-order phase transition regime, one has to rotate $\tilde{r}$ to obtain $r$ satisfying the above condition. 
Since the first-order phase transition line is expected to be bent, it is intuitive to perform this rotation such 
that the $r$-axis lies tangentially to the transition line in the QCD critical point. The exact orientation of the 
$h$-axis is, in contrast, less constrained. In line with~\cite{Bluhm:2016byc}, we opt for defining $h=\tilde{h}$ 
parallel to the $T$-axis in QCD. 

\begin{figure}[t]
\begin{minipage}{0.48\textwidth}
\begin{center}
\includegraphics[width=1\textwidth]{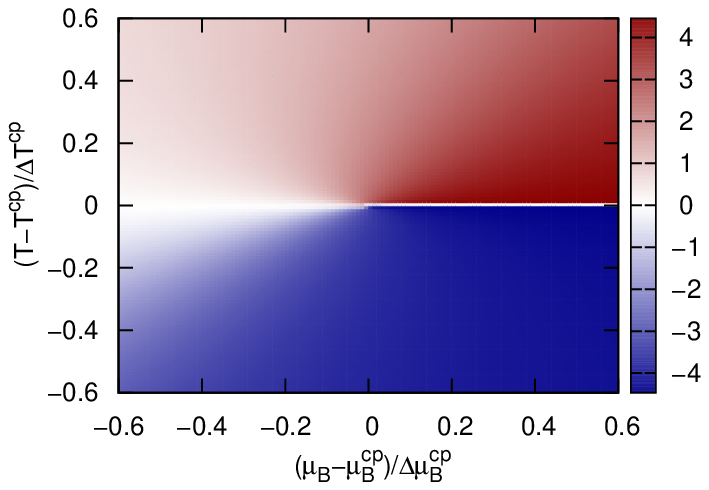}
\caption{\label{Fig:Fig1} Scaled dimensionless parameter $\tilde{\lambda}_3/C$ from Eq.~(\ref{equ:9}) as a 
function of $\tilde{r}$ and $\tilde{h}$, cf.~text for details.}
\end{center}
\end{minipage}\hspace{0.02\textwidth}
\begin{minipage}{0.48\textwidth}
\begin{center}
\includegraphics[width=1\textwidth]{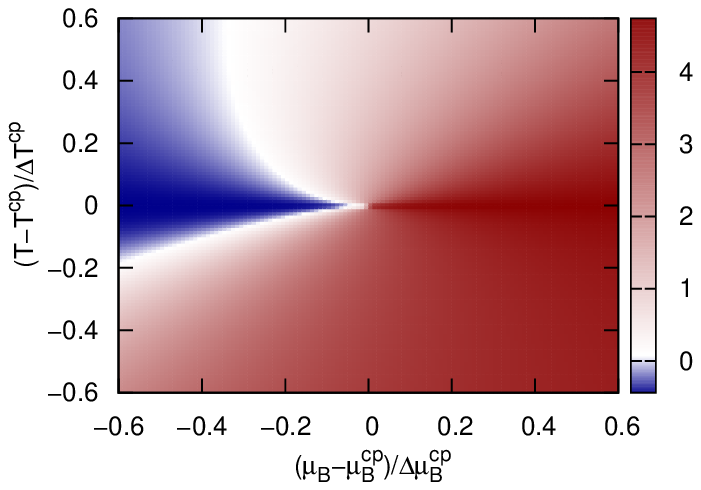}
\caption{\label{Fig:Fig2} As in Figure~\ref{Fig:Fig1} but for the scaled dimensionless parameter difference 
in Eq.~(\ref{equ:10}), $(2\tilde{\lambda}_3^2-\tilde{\lambda}_4)/C^2$.}
\end{center}
\end{minipage}\hfill
\end{figure}
The corresponding behavior of the scaled parameter $\tilde{\lambda}_3/C$ and the scaled parameter difference 
$(2\tilde{\lambda}_3^2-\tilde{\lambda}_4)/C^2$ in the QCD phase diagram is shown in Figures~\ref{Fig:Fig1} 
and~\ref{Fig:Fig2}, respectively. Qualitatively, the observed patterns follow the behavior with $r$ and $h$ 
described in Section~\ref{sec:2}, but appear tilted to some extent given the particular mapping to QCD 
thermodynamics employed in~\cite{Bluhm:2016byc}. 

\section{\label{sec:4} Conclusions}

We discussed the behavior of important parameters of the effective action near the QCD critical point in line with 
the three-dimensional Ising model universality class. The shown results are based on the mapping used 
in~\cite{Bluhm:2016byc}. In future work, these parameters will serve as input for dynamical studies of critical 
fluctuations in heavy-ion collisions, similar to~\cite{Nahrgang:2011mg,Herold:2016uvv}. 

\ack 
The authors thank J.-W.~Chen, S.~Mukherjee and M.~Stephanov for helpful discussions. M.N. acknowledges support 
from a fellowship within the Postdoc-Program of the German Academic Exchange Service (DAAD). This work was 
supported in parts by the U.S. Department of Energy under grants DE-FG02-03ER41260 and DE-FG02-05ER41367, and 
within the framework of the Beam Energy Scan Theory (BEST) Topical Collaboration. 

\section*{References}



\end{document}